\documentclass[aps,prb,longbibliography,reprint,amsmath,amssymb,superscriptaddress]
{revtex4-2}
\usepackage{graphicx}% Include figure files
\usepackage{dcolumn}% Align table columns on decimal point
\usepackage{bm}% bold math
\usepackage{color} % For blue in-text comments and additions
\usepackage{physics}
\usepackage{hyperref}
\usepackage{subfigure}
\hypersetup{
    colorlinks = true,
    linkcolor = blue,
    citecolor = blue,
    anchorcolor = blue,
    urlcolor = blue
    }

\def\lesssim{\ \raise.3ex\hbox{$<$}\kern-0.8em\lower.7ex\hbox{$\sim$}\ }
\def\gesim{\ \raise.3ex\hbox{$>$}\kern-0.8em\lower.7ex\hbox{$\sim$}\ }

\newcommand \beq{\begin{eqnarray}}
\newcommand \eeq{\end{eqnarray}}

\usepackage{amsmath,amssymb}
\usepackage{fixmath}
\usepackage[normalem]{ulem}

\usepackage{comment}

\begin{document}

\preprint{RIKEN-iTHEMS-Report-24}

\title{Probing Goldstino excitation through the tunneling transport in a Bose-Fermi mixture with explicitly broken supersymmetry}

\author{Tingyu Zhang}
\email{zhangty@g.ecc.u-tokyo.ac.jp}
\affiliation{Department of Physics, School of Science, The University of Tokyo, Tokyo 113-0033, Japan}
\affiliation{Interdisciplinary Theoretical and Mathematical Sciences Program (iTHEMS), RIKEN, Wako 351-0198, Japan}

\author{Yixin Guo}
\email{guoyixin1997@g.ecc.u-tokyo.ac.jp}
\affiliation{Department of Physics, School of Science, The University of Tokyo, Tokyo 113-0033, Japan}
\affiliation{Interdisciplinary Theoretical and Mathematical Sciences Program (iTHEMS), RIKEN, Wako 351-0198, Japan}

\author{Hiroyuki Tajima}
\email{htajima@g.ecc.u-tokyo.ac.jp}
\affiliation{Department of Physics, School of Science, The University of Tokyo, Tokyo 113-0033, Japan}

\author{Haozhao Liang}
\email{haozhao.liang@phys.s.u-tokyo.ac.jp}
\affiliation{Department of Physics, School of Science, The University of Tokyo, Tokyo 113-0033, Japan}
\affiliation{Interdisciplinary Theoretical and Mathematical Sciences Program (iTHEMS), RIKEN, Wako 351-0198, Japan}

\begin{abstract}

We theoretically investigate the tunneling transport in a repulsively interacting ultracold Bose-Fermi mixture. 
A two-terminal model is applied to such a mixture
and the supersymmetry-like tunneling current through the junction can be induced by the bias of fermion chemical potential between two reservoirs. 
The goldstino, which is the Nambu-Goldstone fermionic mode associated with the spontaneouls sypersymmetry breaking and appears as a gapped mode in the presence of the explicit supersymmetry breaking in existing Bose-Fermi mixtures, 
is
found to contribute to the tunneling transport as a supercharge exchanging process. 
Our study provides a potential way to detect the goldstino transport in cold atom experiments.
\end{abstract}

\maketitle

\section{Introduction}

Quantum transport phenomena serve as a crucial avenue for understanding complex quantum many-body systems, with extensive exploration in condensed matter physics, atomic physics, and nuclear physics. The unique characteristics of cold atomic systems, such as controllable interaction strength, variable species number, and densities, make them an ideal platform for investigating phenomena challenging to realize in solid-state systems. The Feshbach resonances for adjusting interparticle scattering length enable us to explore strongly interacting regimes~\cite{RevModPhys.82.1225}.
%, facilitating the study of correlated transport phenomena~\cite{PhysRevLett.92.040403,PhysRevLett.92.120401}.
The density manipulation of atomic species allows for the direct induction and measurement of quantum transport~\cite{chien2015quantum,sommer2011universal,pnas.1601812113}. Leveraging the cleanness, high controllability, and advanced experimental techniques of cold atomic systems, 
researchers have observed and investigated diverse transport phenomena, including bulk spin transport~\cite{PhysRevA.88.033630,doi:10.1146/annurev-conmatphys-031218-013732,doi:10.1126/science.1247425}, multiple Andreev reflections~\cite{science.aac9584}, and the Josephson effect~\cite{science350,Krinner_2017,science.aaz2463}.

Recently, an ultracold Bose-Fermi mixture~\cite{epub29268,Ikemachi_2017,PhysRevLett.118.103403,PhysRevA.84.011601,barbe2018observation} 
opens a novel way towards a wide range of applications in investigating quantum many-body systems~\cite{PhysRevLett.90.170403,PhysRevLett.92.050401,PhysRevLett.96.180402,PhysRevA.84.033627,Wang_2020}, including possible realization of the analog quantum simulations toward
supersymmetry~\cite{PhysRevLett.100.090404,PhysRevA.81.011604,PhysRevA.91.063620,PhysRevA.92.063629,PhysRevA.93.033642,PhysRevA.96.063617,PhysRevResearch.3.013035},
dense hadron-quark matter~\cite{PhysRevLett.103.085301,Baym2018Rep.Prog.Phys.81.056902,PhysRevA.103.063317,Guo2023Phys.Rev.A108.023304}
, and neutron-rich nuclei~\cite{Sogo2003,PhysRevA.66.013618,Guo2024PhysRevA109.013319,tajima2024intersections}.
Certain mixtures with small mass imbalance are realized in experiments by using different isotopes, such as $^6$Li-$^7$Li~\cite{doi:10.1126/science.1255380,PhysRevLett.118.103403,Ikemachi_2017}, $^{39}$K-$^{40}$K~\cite{PhysRevA.78.012503}, $^{40}$K-$^{41}$K~\cite{PhysRevA.84.011601}, $^{84}$Sr-$^{87}$Sr~\cite{PhysRevA.82.011608}, $^{161}$Dy-$^{162}$Dy~\cite{PhysRevLett.108.215301}, and $^{173}$Yb-$^{174}$Yb~\cite{PhysRevA.79.021601,sugawa2011interaction}. In such systems, fermionic Nambu-Goldstone mode, which is called the goldstino~\cite{SALAM1974465,WITTEN1981513,Wess:1992cp,PhysRevLett.100.090404}, is generated when the supersymmetry, namely the Fermi-Bose exchange symmetry, is broken. 
While the existence of the bosonic Nambu-Goldstone mode~\cite{PhysRev.124.246,goldstone1961field}
has been confirmed experimentally in a cold atomic system~\cite{hoinka2017goldstone},
the existence of goldstino has been still elusive.
Such a situation is not limited to ultracold atomic physics and it is still difficult to confirm the goldstino in nature.
In this regard, it is an exciting challenge to detect possible appearance of the Goldstino in atomic systems.
\begin{figure}[t]
    \centering
    \includegraphics[width=8.6cm]{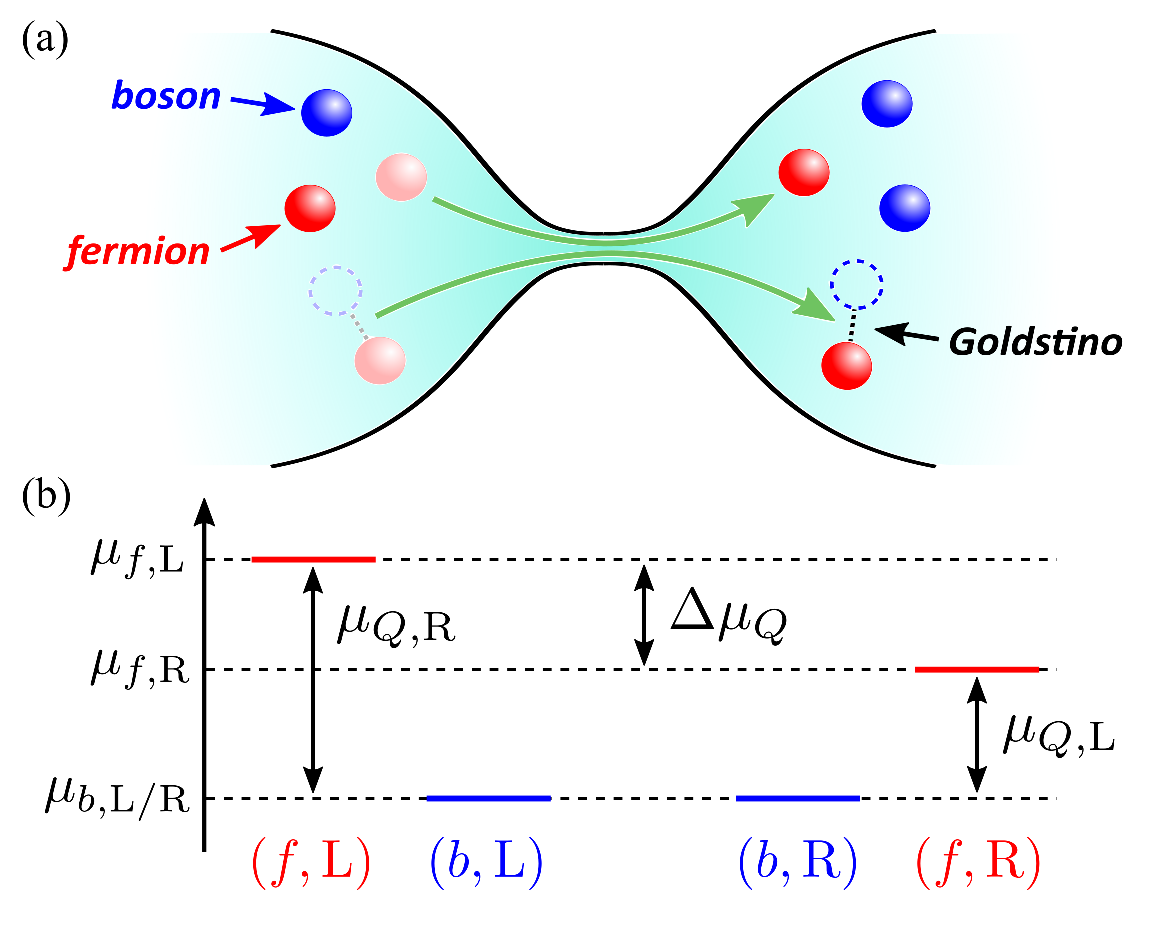}
    \caption{
    (a) Schematic view of the two-terminal system considered in this work. Quasiparticle tunneling and supercharge tunneling are provoked by chemical potential bias. The dashed circle represents the hole, and the pair of a fermion and a bosonic hole denotes a goldstino mode.
    (b) Chemical potentials for bosons (b) and fermions (f) in the left (L) and right (R) reservoirs, where $\mu_{Q,i}$ denotes the chemical potential for supercharge. The bosonic chemical potentials in two reservoirs are set to be equal for suppressing the bosonic quasiparticle and pair tunneling processes.}\label{schematic}
\end{figure}

A hint to find the consequence of goldstino
is that the goldstino can be regarded as an analog of the magnon mode, which is a collective excitation of spin structures in a ferromagnetic phase~\cite{chumak2015magnon} and is also a type of Nambu-Goldstone boson. In repulsively interacting Fermi gases at low temperatures, magnons have been proposed to play a crucial role in spin transport, with its enhanced signal as the interaction strength increases~\cite{PhysRevB.108.155303,PhysRevApplied.21.L031001}. 
While the magnon mode arises from the spin-flip process~\cite{Sandri_2011}, involving the interchange of a spin-up and a spin-down particles, 
the goldstino corresponds to the interchange of a bosonic and a fermionic atoms.
Accordingly, it raises the intriguing question of whether the goldstino contributes to transport in such mixtures and whether it can be identified through the measurement of tunneling current feasible in an ultracold atomic system.
We note that even in the presence of the explicit supersymmetry breaking in a Bose-Fermi mixture (e.g., the imbalance of masses and chemical potentials between fermions and bosons),
one can expect the goldstino excitation but with the nonzero energy gap~\cite{PhysRevResearch.3.013035}.

In this study, we propose a two-terminal model comprising ultracold Bose-Fermi mixtures, allowing for the tuning of species density to induce tunneling processes driven by a chemical potential bias between two reservoirs (see Fig.~\ref{schematic}). The bosons predominantly exist in the Bose-Einstein condensate (BEC) phase with zero momentum below the BEC critical temperature, while the fermions form a fully filled Fermi sea. Assuming the fermions are polarized with only one remaining degree of freedom (i.e., single hyperfine state),
we consider the boson-fermion and boson-boson interactions.
In the case where the mass imbalance between a boson and a fermion is negligible, supersymmetry within each reservoir is explicitly broken by the difference in two interaction strengths and the chemical potential bias between bosons and fermions.
Considering single-particle energy corrections for both bosons in the condensate and normal phases and adopting the Schwinger-Keldysh approach~\cite{Schwinger,Keldysh}, we analyze the supersymmetry-like current, encompassing quasiparticle and goldstino contributions. Using the small mass-balanced mixture
such as $^{173}$Yb-$^{174}$Yb, we present how the goldstino spectrum affects the tunneling transport.

This paper is structured as follows: In Section~\ref{Formalism}, we introduce the Hamiltonian for the total system and the tunneling current operators. 
The formulas for supersymmetry-like tunneling currents are derived up to the leading order in Section~\ref{SUSY current}. 
In Section~\ref{Results}, we explore the spectra and density of states for the goldstino in a given Bose-Fermi mixture and conduct numerical calculations for the supercharge tunneling current with varying chemical potential bias. 
%Our findings are summarized 
Finally, a summary and perspectives are given in Section~\ref{Summary}.

Throughout the paper, we take $\hbar=k_B=1$ and the volumes for both reservoirs to be unity.

\section{Formalism}\label{Formalism}

Here we introduce the two-terminal model for tunneling between two Bose-Fermi mixtures separated by a potential barrier.
The total Hamiltonian of the system is given by
\begin{align}\label{Hamiltonian}
\hat{H}= &
\sum_{\alpha=f,b}\int d^3 \bm{r} \hat{\psi}_\alpha^{\dagger}(\bm{r})\left[-\frac{\nabla^2}{2 m_\alpha}+V_\alpha(\bm{r})\right] \hat{\psi}_\alpha(\bm{r})\cr
& +\frac{U_{b b}}{2} \int d^3 \bm{r} \hat{\psi}_b^{\dagger}(\bm{r}) \hat{\psi}_b^{\dagger}(\bm{r}) \hat{\psi}_b(\bm{r})\hat{\psi}_b(\bm{r}) \cr
& +U_{b f} \int d^3 \bm{r} \hat{\psi}_b^{\dagger}(\bm{r}) \hat{\psi}_b(\bm{r}) \hat{\psi}_f^{\dagger}(\bm{r}) \hat{\psi}_f(\bm{r}),
\end{align}
where $\hat{\psi}_{\alpha}$ is the field operator of a fermion ($\alpha=f$) or a boson ($\alpha=b$).
The $m_{\alpha}$ are the masses and chemical potentials of a fermion and a boson, respectively.
$V_{\alpha}(\bm{r})$ describes the potential barrier separating two reservoirs and goes to zero far away from the junction.
The $U_{bb}$ and $U_{bf}$ are respectively the coupling strengths for the boson-boson and boson-fermion interactions, which are characterized by the scattering lengths $a_{bb}$ and $a_{bf}$ as $U_{bb}=(4\pi a_{bb})/m_b$ and $U_{bf}=(2\pi a_{bf})/m_r$ with $m_r=1/(1/m_b+1/m_f)$ denoting the reduced mass.

For the steady-state transport between two reservoirs,
the field operator can be decomposed as~\cite{PhysRevA.106.033310} 
\begin{align}\label{decomposition}
    \hat{\psi}_{\alpha}(\bm{r})=\hat{\psi}_{\alpha,{\rm L}}(\bm{r})+\hat{\psi}_{\alpha,{\rm R}}(\bm{r}),
\end{align}
where $\hat{\psi}_{\alpha,i}(\bm{r})$ denotes the field operator in the reservoir $i={\rm L, R}$. 
While the potential barrier peaking in the junction between the reservoirs induce an inhomogeneity near the barrier, in the far region the potential goes smoothly to zero. We can therefore consider uniform gases inside the reservoirs, with the wave function being the asymptotic form:
\begin{align}\label{psi_l}
{\psi}_{\bm{k},\alpha, {\rm L}}(\bm{r})= 
\begin{cases}
e^{i \bm{k} \cdot \bm{r}}+R_{\bm{k}, \alpha} e^{-i \bm{k} \cdot \bm{r}} & (x<0) \\
T_{\bm{k}, \alpha} e^{i \bm{k} \cdot \bm{r}} & (x>0),
\end{cases}
\end{align}
\begin{align}\label{psi_r}
{\psi}_{\bm{k},\alpha, {\rm R}}(\bm{r})=
%\sum_{\bm{k}}
%\widetilde{\alpha}_{\bm{k},{\rm R}}\times 
\begin{cases}
T_{\bm{k}, \alpha} e^{-i \bm{k} \cdot \bm{r}} & (x<0), \\
e^{-i \bm{k} \cdot \bm{r}}+R_{\bm{k}, \alpha} e^{i \boldsymbol{k} \cdot \boldsymbol{r}} & (x>0),
\end{cases}
\end{align}
where $\bm{k}$ is the wave number.
%$\widetilde{\alpha}_{\bm{k},i}$ is the amplitude of the asymptotic wave function of a boson ($\widetilde{\alpha}=\widetilde{b}$)
%or a fermion ($\widetilde{\alpha}=\widetilde{f}$). 
The potential scattering induces
the one-particle reflection and transmission coefficients
$\mathcal{R}_{\bm{k},\alpha}$
 and $\mathcal{T}_{\bm{k},\alpha}$, respectively.
In Eqs.~\eqref{psi_l} and \eqref{psi_r}, the coordinate $x$ symbolically denotes the direction perpendicular to the potential barrier located at $x=0$. 
Using Eqs.~\eqref{psi_l} and \eqref{psi_r}, we expand $\hat{\psi}_{\alpha,i}(\bm{r})$ in Eq.~\eqref{decomposition} as
\begin{align}
\label{eq:5}
    \hat{\psi}_{f,i}(\bm{r})=\sum_{\bm{k}}\psi_{\bm{k},f,i}(\bm{r})\hat{f}_{\bm{k},i},
    \\
    \hat{\psi}_{b,i}(\bm{r})=\sum_{\bm{k}}\psi_{\bm{k},b,i}(\bm{r})\hat{b}_{\bm{k},i},
    \label{eq:6}
\end{align}
where $\hat{f}_{\bm{k},i}$ and $\hat{b}_{\bm{k},i}$ are annihilation operators of a fermion and a boson, respectively.
By substituting Eqs.~\eqref{eq:5} and \eqref{eq:6} into Eq.~(\ref{decomposition}) and then Eq.~(\ref{Hamiltonian}),
%and replacing $\widetilde{\alpha}_{\bm{k},i}$ with the annihilation operator $\alpha_{\bm{k},i}$
we obtain the Hamiltonian of the system as 
\begin{align}
    \hat{H}=\hat{H}_{\rm L}+\hat{H}_{\rm R}+\hat{H}_{\rm 1t}+\hat{H}_{\rm 2t},
\end{align}
where the reservoir Hamiltonian reads 
\begin{align}
\hat{H}_{i={\rm L, R}}=\,&
\sum_{\bm{k}} 
\varepsilon_{\bm{k}, f} 
\hat{f}_{\bm{k}, i}^{\dagger} 
\hat{f}_{\bm{k}, i} 
+\sum_{\bm{k}} 
\varepsilon_{\bm{k}, b} 
\hat{b}_{\bm{k}, i}^{\dagger} \hat{b}_{\bm{k}, i} \cr
&+\frac{U_{bb}}{2} \sum_{\bm{P}, \bm{q}, \bm{q}'} \hat{b}_{\frac{\bm{P}}{2}+\bm{q}, i}^{\dagger} 
\hat{b}_{\frac{\bm{P}}{2}-\bm{q}, i}^{\dagger} 
\hat{b}_{\frac{\bm{P}}{2}-\bm{q}', i} 
\hat{b}_{\frac{\bm{P}}{2}+\bm{q}', i}\cr
&+U_{bf} \sum_{\bm{P}, \bm{q}, \bm{q}'} \hat{b}_{\frac{\bm{P}}{2}+\bm{q}, i}^{\dagger} 
\hat{f}_{\frac{\bm{P}}{2}-\bm{q}, i}^{\dagger} 
\hat{f}_{\frac{\bm{P}}{2}-\bm{q}', i} 
\hat{b}_{\frac{\bm{P}}{2}+\bm{q}', i},
\end{align}
with the single-particle energy defined as $\varepsilon_{\bm{p},\alpha=b,f}=p^2/(2m_{\alpha})$. This Hamiltonian includes the single-particle terms (the first line of the right hand side), the boson-boson interaction term (the second line), and the boson-fermion interaction term (the third line). The one-body tunneling Hamiltonian yields
\begin{align}
    \hat{H}_{\rm 1t}=&\sum_{\bm{k}_1,\bm{k}_2}\mathcal{T}_{\bm{k}_1,\bm{k}_2,f}\big[\hat{f}^\dagger_{\bm{k}_1,{\rm L}}\hat{f}_{\bm{k}_2,{\rm R}}+\hat{f}^\dagger_{\bm{k}_1,{\rm R}}\hat{f}_{\bm{k}_2,{\rm L}}\big]\cr
    &+\sum_{\bm{k}_1,\bm{k}_2}\mathcal{T}_{\bm{k}_1,\bm{k}_2,b}\big[\hat{b}^\dagger_{\bm{k}_1,{\rm L}}\hat{b}_{\bm{k}_2,{\rm R}}+\hat{b}^\dagger_{\bm{k}_1,{\rm R}}\hat{b}_{\bm{k}_2,{\rm L}}\big],
\end{align}
which is associated with the potential barrier. The one-body tunneling amplitudes are given by
\begin{subequations}
    \begin{align}
        \mathcal{T}_{\bm{k}_1,\bm{k}_2,f}=\mathcal{C}_{\bm{k}_1, \bm{k}_2, f}\bigg[&\delta_{\bm{k}_1, \bm{k}_2}\varepsilon_{\bm{k}_1, f}
        +V_f(\bm{k}_1-\bm{k}_2)\notag\\
        &+\frac{U_{fb}}{2}\sum_i \hat{N}_{b,\bm{k}_1-\bm{k}_2,i}\bigg]\\
        \mathcal{T}_{\bm{k}_1,\bm{k}_2,b}=\mathcal{C}_{\bm{k}_1, \bm{k}_2, b}\bigg[&\delta_{\bm{k}_1, \bm{k}_2}\varepsilon_{\bm{k}_1, b}+V_b(\bm{k}_1-\bm{k}_2)\notag\\
        +\frac{U_{bb}}{2}\sum_i \hat{N}&_{b,\bm{k}_1-\bm{k}_2,i}+\frac{U_{bf}}{2}\sum_i \hat{N}_{f,\bm{k}_1-\bm{k}_2,i}\bigg],
    \end{align}
\end{subequations}
where $\hat{N}_{\alpha,\bm{k},i}$ is the density operator in the reservoir $i$. We define $\mathcal{C}_{\bm{k}_1, \bm{k}_2, \alpha}=\int\,\textrm{d}^3\bm{r}\,\psi^\ast_{\bm{k}_1, \alpha,{\rm L}}(\bm{r})\psi_{\bm{k}_2, \alpha, {\rm R}}(\bm{r})$ as the overlap integral of wave functions between two reservoirs, which is proportional to the transmission coefficient $T_{\bm{k},\alpha}$ in Eqs.~(\ref{psi_l}) and (\ref{psi_r}).
Notice that the $V_{\alpha}(\bm{k})$ is the Fourier transformed potential barrier, which yields a constant $V_{0,\alpha}$ as we approximately use a delta function for the barrier $V_{\alpha}(x)=V_{0,\alpha}\delta(x/\lambda)$, where $\lambda$ is a typical length scale of the barrier width.

The two-body tunneling Hamiltnoian can be written as the sum of three terms $\hat{H}_{\rm 2t}=\hat{H}_{bb}+\hat{H}_{bf}+\hat{H}_{Q}$, respectively corresponding to the tunneling of boson-boson pairs, boson-fermion  pairs, and supercharges. The pair tunneling terms read 
\begin{subequations}
    \begin{align}
        \hat{H}_{bb}=\,&\frac{1}{2}\mathcal{G}_{bb}\sum_{\bm{p},\bm{q}}\big[\hat{P}^\dagger_{bb, {\rm L}}(\bm{p})\hat{P}_{bb, {\rm R}}(\bm{q})+\textrm{h.c.}\big],
    \end{align}
    \begin{align}
        \hat{H}_{bf}=\,&\mathcal{G}_{bf}\sum_{\bm{p},\bm{q}} \big[\hat{P}^\dagger_{bf, {\rm L}}(\bm{p})\hat{P}_{bf, {\rm R}}(\bm{q})+\textrm{h.c.}\big],
    \end{align}
\end{subequations}
where $\hat{P}_{bb,i}$ and $\hat{P}_{bf,i}$ are respectively the annihilation operators for a boson-boson pair and a boson-fermion pair, while $\mathcal{G}_{bb}$ and $\mathcal{G}_{bf}$ are the pair tunneling amplitude.

Now we introduce the supercharge operator $\hat{Q}$ and $\hat{Q}^\dagger$, which are generators of supersymmetry transformations~\cite{bilal2001introduction}. They can be defined in the second quantized notation as 
\begin{align}
    \hat{Q}_i(\bm{p})=\sum_{\bm{k}}\hat{f}_{\bm{k},i}\hat{b}^\dagger_{\bm{k}-\bm{p},i}\ ,\quad 
    \hat{Q}^\dagger_i(\bm{p})=\sum_{\bm{k}}\hat{b}_{\bm{k}-\bm{p},i}\hat{f}^\dagger_{\bm{k},i}.
\end{align}
The supercharge operators satisfy the following relations: $\hat{Q}_i^2=(\hat{Q}_i^\dagger)^2=0$, $\{\hat{Q}_i,\hat{Q}_i^\dagger\}=\hat{N}_i=\hat{N}_{b,i}+\hat{N}_{f,i}$, $[\hat{Q}_i,\hat{N}_i]=[\hat{Q}_i^\dagger,\hat{N}_i]=0$, and thus they are fermionic operators.
Here $\hat{N}_{f,i}=\sum_{\bm{p}}\hat{f}^\dagger_{\bm{p},i}\hat{f}_{\bm{p},i}$ and $\hat{N}_{b,i}=\sum_{\bm{p}}\hat{b}^\dagger_{\bm{p},i}\hat{b}_{\bm{p},i}$ are the particle number of fermions and bosons in the reservoir $i$.
Physically, $\hat{Q}$ changes a fermion into a boson, while $\hat{Q}^\dagger$ does the opposite. The supercharge tunneling term can then be written as
\begin{align}
    {H}_{Q}=\,\mathcal{G}_{Q}\sum_{\bm{p},\bm{q}}\left[\hat{Q}_{\rm L}^\dagger(\bm{p})\hat{Q}_{\rm R}(\bm{q})+\textrm{h.c.}
\right],
\end{align}
with the tunneling amplitude $\mathcal{G}_{Q}$.
Taking the long-wavelength limit for the transmitted waves, the amplitudes of the two-body tunneling terms are given by $\mathcal{G}_{bb}=U_{bb}\operatorname{Re}[T_{\bm{0},b}^2]$, $\mathcal{G}_{bf}=U_{bf}\operatorname{Re}[T_{\bm{0},b}T_{\bm{0},f}]$, and $\mathcal{G}_{Q}=U_{bf}\operatorname{Re}[T_{\bm{0},b}T^*_{\bm{0},f}]$~\cite{10.1093/pnasnexus/pgad045,PhysRevA.106.033310}. Notice that we consider the effective Hamiltonian for the system where we omit the particle reflection and the induced interface interaction, which are irrelevant to our study. 

The supersymmetry-like current operator is given by 
\begin{align}
    \hat{I}_{\rm SUSY}=i\big[\hat{N}_{b,{\rm L}}-\hat{N}_{f,{\rm L}},\hat{H}\big].
\end{align}
It can be rewritten as $\hat{I}_{\rm SUSY}=\hat{I}_{\rm 1t}+\hat{I}_{bb}+\hat{I}_{bf}+\hat{I}_{Q}$
with the quasiparticle tunneling, boson-boson pair tunneling, boson-fermion pair tunneling, and supercharge tunneling contributions as
\begin{align}
    \hat{I}_{\rm 1t}=\,&i\sum_{\bm{p}, \bm{q}} \Big[\mathcal{T}_f\big(\hat{f}_{\bm{p},  {\rm R}}^{\dagger} \hat{f}_{\bm{q}, {\rm L}}-\hat{f}^\dagger_{\bm{p}, {\rm L}}\hat{f}_{\bm{q}, {\rm R}}\big)\cr
    &+\mathcal{T}_b\big(\hat{b}^\dagger_{\bm{p}, {\rm L}}\hat{b}_{\bm{q}, {\rm R}}-\hat{b}_{\bm{p},  {\rm R}}^{\dagger} \hat{b}_{\bm{q}, {\rm L}}\big)\Big],
\end{align}
\begin{align}
    \hat{I}_{bb}=&\,i\mathcal{G}_{bb}\sum_{\bm{p},\bm{q}}\hat{P}^\dagger_{bb, {\rm L}}(\bm{p})\hat{P}_{bb, {\rm R}}(\bm{q})+\textrm{h.c.},
\end{align}
\begin{align}
    \hat{I}_{bf}=\,2i&\mathcal{G}_{bf}\sum_{\bm{p},\bm{q}} \hat{P}^\dagger_{bf, {\rm L}}(\bm{p})\hat{P}_{bf, {\rm R}}(\bm{q})+\textrm{h.c.},
\end{align}
\begin{align}
    \hat{I}_{Q}=\,2i&\mathcal{G}_{Q}\sum_{\bm{p},\bm{q}}\hat{Q}_{\rm L}^\dagger(\bm{p})\hat{Q}_{\rm R}(\bm{q})+\textrm{h.c.},
\end{align}
respectively.
Here, we approximately use the momentum-independent one-body tunneling amplitudes $\mathcal{T}_f=\mathcal{C}_{0,0,f}(\epsilon_{\rm F}+V_{0,f})$ and $\mathcal{T}_b=\mathcal{C}_{0,0,b}(\epsilon_{b}+V_{0,b})$, which correspond to the momentum-conserved tunnelings near the Fermi energy $\epsilon_{\rm F}=(6\pi^2N_f)^{2/3}/(2m_f)$ and the bosonic energy scale $\epsilon_b=(6\pi^2N_b)^{2/3}/(2m_b)$, with the transmission coefficient at long-wavelength limit ($\bm{k}\rightarrow 0$).
$N_{\alpha}=\langle \hat{N}_{\alpha,{\rm L}}\rangle$ is the statistical average of the particle numbers in the reservoir ${\rm L}$.
A similar two-terminal model has been applied to study the mass current and spin current in strongly correlated Fermi gases~\cite{PhysRevA.106.033310,PhysRevB.108.155303,PhysRevResearch.2.023152}.  
On the one hand, the supercharge tunneling can be regarded as an analog of the spin-flip process if we replace the supercharge operators with spin ladder operators.
Physically, the interchange of a fermion and a boson corresponds to the interchange of spin-up and spin-down particles in a two-component Fermi gas.
On the other hand, the spin-ladder operators involve the bosonic commutation relations in contrast to the fermionic supercharge operator $Q$. Such a difference of the operator properties leads to the different distribution functions in the tunneling current.

\section{Supersymmetry-like current}\label{SUSY current}

Using the current operator in the Heisenberg picture, in this section we derive the formulas of the supersymmetry-like currents by applying the Schwinger-Keldysh approach. Taking the tunneling Hmiltonian as a perturbation term, the expectation value of current can be expanded as 
\begin{align}
    I_{\rm SUSY}(t,t')=&\sum_{n=0}^{\infty}\frac{(-i)^n}{n!}\int_Cdt_1\cdots\int_Cdt_n\nonumber\\
    &\langle {\rm T}_C \hat{I}_{\rm SUSY}(t,t')\hat{H}_{\rm t}(t_1)\cdots \hat{H}_{\rm t}(t_n)\rangle,
\end{align}
where $\hat{H}_{\rm t}=\hat{H}_{\rm 1t}+\hat{H}_{\rm 2t}$. The time integral is performed over the Keldysh contour $C$, comprising both a backward and a forward branch, with ${\rm T}_C$ denoting the contour-time-ordering product. The time arguments on different branches are distinguished by the notations $t$ and $t'$.
It is important to note that the Schwinger-Keldysh formalism is well-suited for describing nonequilibrium systems, where operators evolve in the interaction picture with the Hamiltonian $\hat{H}_0 = \sum_{\bm{k},i}\varepsilon_{\bm{k},f} \hat{f}_{\bm{k}, i}^{\dagger} \hat{f}_{\bm{k}, i} + \sum_{\bm{k},i}\varepsilon_{\bm{k},b} \hat{b}_{\bm{k}, i}^{\dagger} \hat{b}_{\bm{k}, i}$. 
We assume local equilibrium within each reservoir described by the chemical potential $\mu_{\alpha,i}$, where operators evolve according to the grand-canonical Hamiltonian $\hat{K}_0 = \hat{H}_0-\mu_{f,i}\hat{N}_{f,i}-\mu_{b,i}\hat{N}_{b,i}$. Consequently, we perform a transformation on the creation and annihilation operators: $\hat{f}^\dagger_{\bm{k},i}(t)\rightarrow e^{i\mu_{f,i}t} \hat{f}^\dagger_{\bm{k},i}(t)$, $\hat{f}_{\bm{k},i}(t)\rightarrow e^{-i\mu_{f,i}t} \hat{f}_{\bm{k},i}(t)$, $\hat{b}^\dagger_{\bm{k},i}(t)\rightarrow e^{i\mu_{b,i}t} \hat{b}^\dagger_{\bm{k},i}(t)$, and $\hat{b}_{\bm{k},i}(t)\rightarrow e^{-i\mu_{b,i}t} \hat{b}_{\bm{k},i}(t)$.

Applying the Langreth rule to convert the integral over the Keldysh contour to that over the real-time axis and subsequently performing the Fourier transform, we derive the expressions for the quasiparticle and pair tunneling contributions to the supersymmetry-like current while retaining the truncation at the leading order,
\begin{widetext}
    \begin{align}\label{I1t1}
        I_{\rm 1t}=&\,4 \mathcal{T}_f^2 \int \frac{d \omega}{2 \pi} \sum_{\bm{p}, \bm{q}}
        %\mathcal{T}_{\bm{p}, \bm{q}, f} \mathcal{T}_{\bm{q}, \bm{p}, f}
        \operatorname{Im} G_{f, \bm{p}, \mathrm{L}}(\omega-\Delta \mu_f) \operatorname{Im} G_{f, \bm{q}, \mathrm{R}}(\omega)\left[f_{f}(\omega-\Delta \mu_f)-f_{f}(\omega)\right]\nonumber\\
        &-4
        \mathcal{T}_b^2
        \int \frac{d \omega}{2 \pi} \sum_{\bm{p}, \bm{q}}
        %\mathcal{T}_{\bm{p}, \bm{q}, b} \mathcal{T}_{\bm{q}, \bm{p}, b}
        \operatorname{Im} G_{b, \bm{p}, \mathrm{L}}(\omega-\Delta \mu_b) \operatorname{Im} G_{b, \bm{q}, \mathrm{R}}(\omega)\left[f_{b}(\omega-\Delta \mu_b)-f_{b}(\omega)\right],
    \end{align}
    \begin{align}
        I_{bb}=2 \mathcal{G}_{bb}^{2} \sum_{\bm{p},\bm{q}} \int \frac{d \Omega}{2 \pi} \operatorname{Im} \Gamma_{bb, \bm{p}, {\rm L}}(\Omega-2\Delta\mu_b) \operatorname{Im} \Gamma_{bb, \bm{q}, \mathrm{R}}(\Omega)\left[f_b(\Omega-2\Delta\mu_b)-f_b(\Omega)\right],
    \end{align}
    \begin{align}
        I_{bf}=8\mathcal{G}_{bf}^{2}\sum_{\bm{p},\bm{q}} \int \frac{d \Omega}{2 \pi} \operatorname{Im} \Gamma_{bf, \bm{p}, {\rm L}} (\Omega-(\Delta\mu_f+\Delta\mu_b))\operatorname{Im} \Gamma_{bf, \bm{q}, \mathrm{R}}(\Omega)\left[f_f(\Omega-(\Delta\mu_f+\Delta\mu_b))-f_f(\Omega)\right].
    \end{align}
\end{widetext}
Here, $G_{\alpha,\bm{p},i}(\omega)$ and $\Gamma_{bb(bf),\bm{p},i}(\Omega)$ are respectively the Fourier decompositions of the retarded Green's functions for single particle and pair, defined as $G_{\alpha,\bm{p},i}(t,t')=-i\theta(t-t')\langle\{\hat{\alpha}_{\bm{p},i}(t)\hat\alpha^\dagger_{\bm{p},i}(t')\}\rangle$ ($\hat{\alpha}=\hat{f},\hat{b}$), $\Gamma_{bb,\bm{p},i}(t,t')=-i\theta(t-t')\langle[\hat{P}_{bb,\bm{p},i}(t)\hat{P}^\dagger_{bb,\bm{p},i}(t')]\rangle$, and $\Gamma_{bf,\bm{p},i}(t,t')=-i\theta(t-t')\langle\{\hat{P}_{bf,\bm{p},i}(t)\hat{P}^\dagger_{bf,\bm{p},i}(t')\}\rangle$ with brackets $[\cdots]$ and $\{\cdots\}$ denoting the commutation and anti-commutation operations. The chemical potential bias is denoted as $\Delta\mu_{\alpha}=\mu_{\alpha,{\rm L}}-\mu_{\alpha,{\rm R}}$ and $f_{\alpha}(\omega)$ is the distribution function
%$f_{f}(\omega)=1/(e^{\omega/T}+ 1)$ 
obtained from the relations between the lesser and retareded Green's functions: $G^{<}_{\alpha}(\omega)=\mp2 i \operatorname{Im} G_{\alpha}(\omega) f_{\alpha}(\omega)$ ($-$ for $\alpha=f$ and $+$ for $\alpha=b$)
and $\Gamma_{bb(bf)}^{<}(\Omega)=\pm2 i \operatorname{Im} \Gamma_{bb(bf)}(\Omega) f_{b(f)}(\Omega)$.
\begin{figure}[t]
    \centering
    \includegraphics[width=8.6cm]{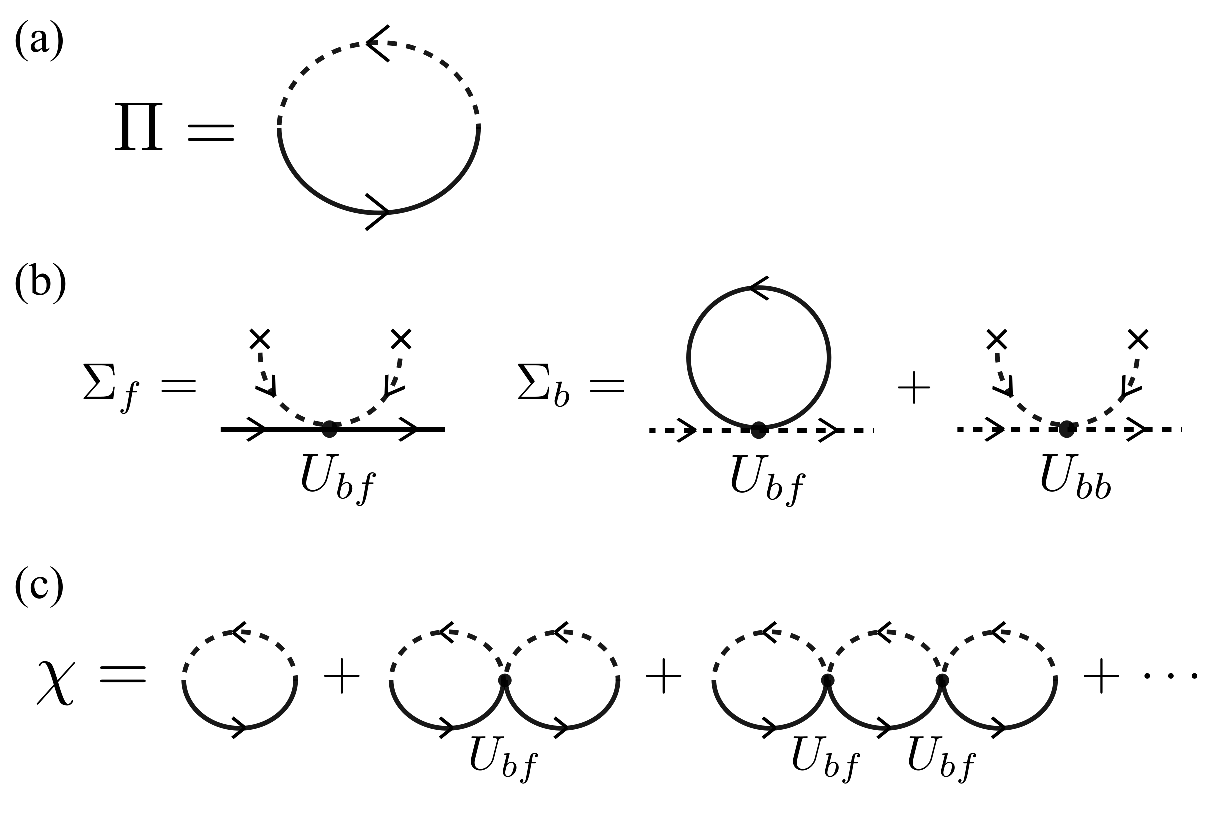}
    \caption{(a) One-loop diagram for noninteracting goldstino propagator. The solid (dashed) line represents the fermion (boson) propagator. (b) Diagrams of leading-order self-energies of fermions (the first one) and bosons (the second and third ones). The cross connected to one end of a dahsed line denotes the contribution of the boson's condensate.
    (c) RPA calculations for the explicit propagator of goldstino. 
    }
    \label{diagrams}
\end{figure} 

We then introduce the goldstino propagator, which is defined as $\chi_{\bm{p}}(t,t')=i\theta(t-t')\langle\{\hat{Q}_{\bm{p}}(t),\hat{Q}^\dagger_{\bm{p}}(t')\}\rangle$ in the linear response theory. The relation between the lesser component and retarded one reads $\chi^{<}(\Omega)=-2i\operatorname{Im} \chi(\Omega)f_f(\Omega)$. Defining $\mu_{Q,i}=\mu_{f,i}-\mu_{b,i}$, the goldstino tunneling current can then be written as
\begin{align}\label{Isc}
    I_{Q}=8\mathcal{G}_{Q}^{2} \sum_{\bm{p},\bm{q}} \int \frac{d \Omega}{2 \pi} &\operatorname{Im} \chi_{\bm{p}, {\rm L}}(\Omega-\Delta\mu_{Q}) \operatorname{Im} \chi_{\bm{q}, {\rm R}}(\Omega)\nonumber\\
    \times&\left[f_f(\Omega-\Delta\mu_{Q})-f_f(\Omega)\right],
\end{align}
with $\Delta\mu_{Q}=\mu_{Q,L}-\mu_{Q,R}$. To induce the supercharge tunneling current with $I_{\rm pair}$ suppressed, we shall take $\mu_{b,{\rm L}}=\mu_{b,{\rm R}}$ but different fermion chemical potentials in the two reservoirs.

\section{Numerical results}\label{Results}

In order to search for evidence of the existence of the goldstino, in the following calculations, we concentrate on the supercharge tunneling current, neglecting the boson-fermion pair tunneling, which should be suppressed by the repulsive interaction between a boson and a fermion. It is noted that the repulsive quantum gas encounters an instability towards pair foramtion~\cite{Massignan_2014,PhysRevLett.129.203402}, which would compete with a phase separation in strongly interacting regime~\cite{PhysRevLett.121.253602}. Nevertheless, to avoid this we consider the weakly interacting case, where the two-particle boundary state is so deep and the Bose-Fermi mixture remains at a metastable equilibrium state, such that our approach is valid.

\subsection{Spectral functions in bulk reservoirs}

First, we calculate the spectral functions in bulk reservoirs and thus we suppress the indices for the reservoirs ($i={\rm L,R}$) for convenience in this subsection. 
The one-loop diagram of the bare goldstino propagator $\Pi$ includes a fermion propagator and a boson propagator as drawn in Fig.~\ref{diagrams}(a). The explicit form of $\Pi_{\bm{p}}(\Omega)$ is given by
\begin{align}\label{Lindhard}
    \Pi_{\bm{p}}(\Omega)=-\int\frac{d^3\bm{k}}{(2\pi)^3}
    \frac{f_f(\xi_{\bm{k}+\bm{p},f})+f_b(\xi_{\bm{k},b})}{\Omega+i\delta-\xi_{\bm{k}+\bm{p},f}+\xi_{\bm{k},b}}.
\end{align}
Here we define $\xi_{\bm{k},f}=\bm{k}^2/(2m_f)-\mu_f+\Sigma_{f}$ and $\xi_{\bm{k},b}=\bm{k}^2/(2m_b)-\mu_b+\Sigma_{b}$, where $\Sigma_f$ and $\Sigma_b$ denote the mean-field corrections for the single-particle energies of fermions and bosons.
We consider the leading-order correction caused by the self-energies below the BEC critical temperature as shown in Fig.~\ref{diagrams}(b), where the first diagram describes that of a fermion while the last two describe that of a boson. 
For fermions and bosons with nonzero momenta, 
the self-energies read
$\Sigma_f=U_{bf}N_b$ and $\Sigma_{b}=2U_{bb}N_b+U_{bf}N_f$, respectively.
% where $N_{\alpha}=\langle \hat{N}_{\alpha}\rangle$ is the statistical average of the particle numbers.
For condensed bosons, it turns out to be $\Sigma_{b,0}=U_{bb}N_b+U_{bf}N_f$ due to the absence of an exchange term in the former case~\cite{PhysRevA.96.063617}. 
The chemical potentials for bosons and fermions can be given as $\mu_b=d E/d N_b=U_{bb}N_b+U_{bf}N_f$ and $\mu_f=d E/d N_f=\epsilon_{\rm F}+U_{bf}N_b$.
Accordingly, the Hugenholtz-Pines condition~\cite{RevModPhys.76.599}, that is, the gapless condition in the condensed phase is fulfilled as $\xi_{\bm{k}=\bm{0},b}\equiv-\mu_b+\Sigma_{b,0}=0$.
However, for nonzero $\bm{k}$, the dispersion reads
$\xi_{\bm{k},b}=k^2/(2m_b)+U_{bb}N_b$~\cite{pethick2008bose} because the present approximation ignores the coupling with the hole-like excitation given by $\hat{b}_{\bm{k}}^\dag \hat{b}_{-\bm{k}}^\dag$ and $\hat{b}_{-\bm{k}}\hat{b}_{\bm{k}}$, which can be justified at $|\bm{k}|\gesim \sqrt{2mU_{bb}N_b}$~\cite{PhysRevA.96.063617}.
Notice that apart from the mean-field approach we applied in this paper, the quasiparticle spectrums can be modified by beyond-mean-field theories~\cite{shi1998finite}. In this regard, we consider the weak-coupling regime where such corrections are small.
%such as the Hugenholtz-Pines theorem~\cite{ENOMOTO20061892,PhysRevA.103.053307}, which is commonly employed in standard bosonic theories such as Bogoliubov theory.}

Taking the mass-balanced case $m_b=m_f=m$, we can rewrite the one-loop propagator as 
\begin{align}\label{Lindhard1}
    \Pi_{\bm{p}}(\Omega)=&-\frac{N^0_b}{\Omega+\epsilon_{\rm F}+i\delta-\bm{p}^2/2m}\cr
    &-\int\frac{d^3\bm{k}}{(2\pi)^3}
    \frac{{f}_f(\xi_{\bm{k}+\bm{p},f})+{f}_b(\xi_{\bm{k},b})}{\Omega+i\delta-\xi_{\bm{k}+\bm{p},f}+\xi_{\bm{k},b}}\cr
    =&~\Pi^{\rm p}_{\bm{p}}(\Omega)+\Pi^{\rm c}_{\bm{p}}(\Omega),
\end{align}
where $N^0_b$ is the number of bosons in the condensate, while $\xi_{\bm{k},f}=\bm{k}^2/2m-\epsilon_{\rm F}$ and $\xi_{\bm{k},b}=\bm{k}^2/2m+U_{bb} N_b$. $\Pi^{\rm p}$ and $\Pi^{\rm c}$ respectively denote the contribution to the spectrum from the pole and continuum. When the temperature approaches zero, $N^0_b\rightarrow N_b$ while the distribution of bosons with $k>0$ vanishes, and 
\begin{align}\label{Pic0}
    \Pi^{\rm c}_{\bm{p}}(\Omega)\rightarrow -\int\frac{d^3\bm{k}}{(2\pi)^3}
    \frac{{f}_f(\xi_{\bm{k}+\bm{p},f})}{\Omega+i\delta-\xi_{\bm{k}+\bm{p},f}+\xi_{\bm{k},b}}.
\end{align}

\begin{figure}[t]
    \centering
    \includegraphics[width=8.6cm]{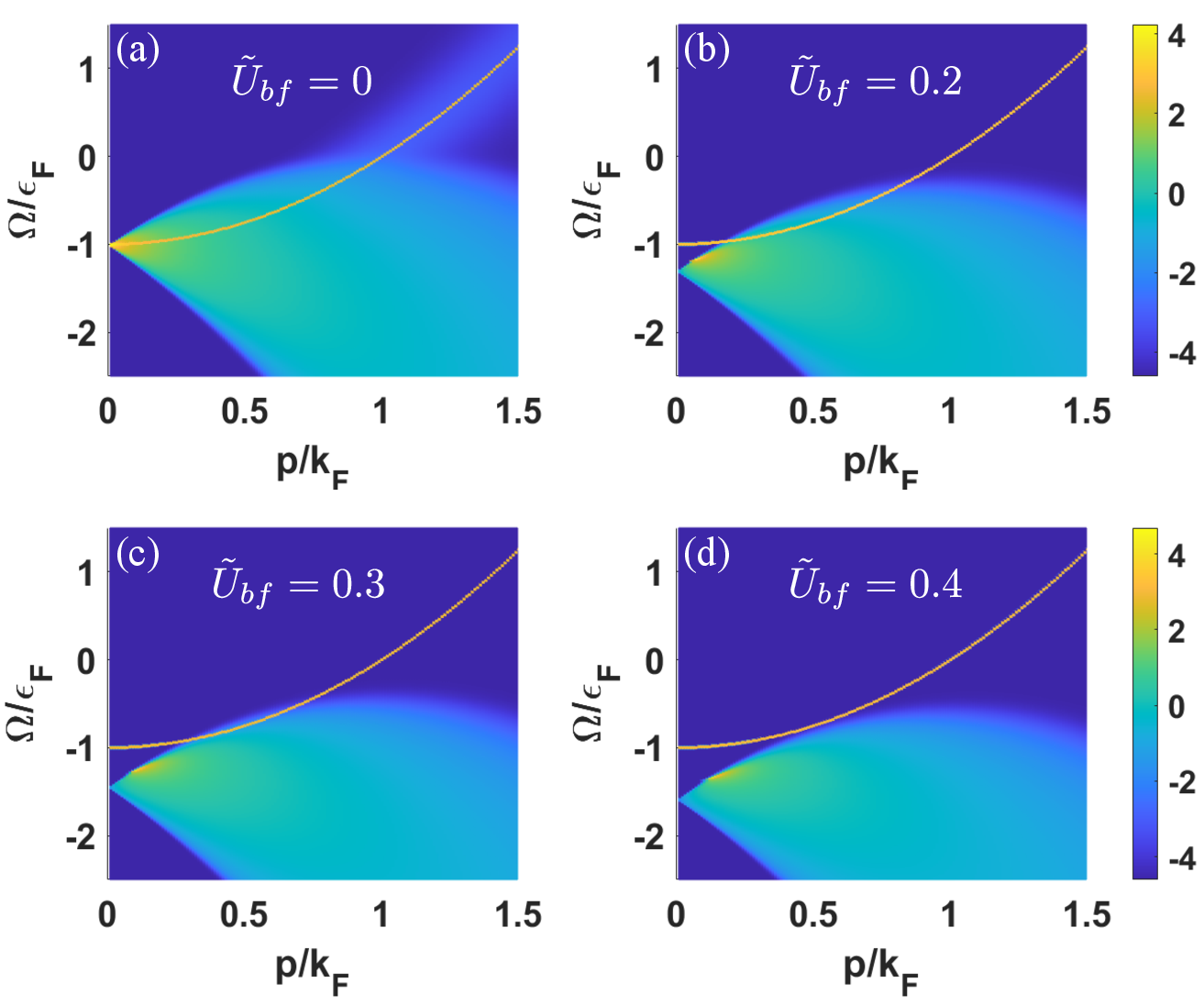}
    \caption{
    goldstino spectra with different particle densities in a Bose-Fermi mixture with explicitly broken supersymmetry. The color bars show the logarithmic scale. We fix the ratio $N_b/N_f=2$ and the interaction parameters are chosen to be the experimental values of the $^{173}$Yb-$^{174}$Yb mixture, $a_{bf}/a_{bb}\simeq1.32$. The goldstino poles (bright yellow line) originates from the BEC phase and the continuum (light area) does from the excitation of bosons. As density becomes larger, the poles start to separate from the continuum. 
    }\label{Imchi}
\end{figure}

We then use the random phase approximation (RPA) to calculate the explicit result of goldstino propagator in the interacting regime.
From Eq.~(\ref{Lindhard}), we can see that the one-loop diagram for $\Pi$ has the order of magnitude $U^{-1}$, with $U$ denoting the magnitude of interaction strengths. We can draw a series of diagrams consisting of $\Pi$ as shown in Fig.~\ref{diagrams}(c), which have the same order of magnitude $U^{-1}$, as one loop contributes $U^{-1}$ and one dot contributes $U$. Thus all of these diagrams should be summed, yielding 
\begin{align}\label{RPA}
    \chi_{\bm{p}}(\Omega)=\frac{\Pi_{\bm{p}}(\Omega)}{1+U_{bf}\Pi_{\bm{p}}(\Omega)}.
\end{align}
\begin{figure}[t]
    \centering
    \includegraphics[width=8.6cm]{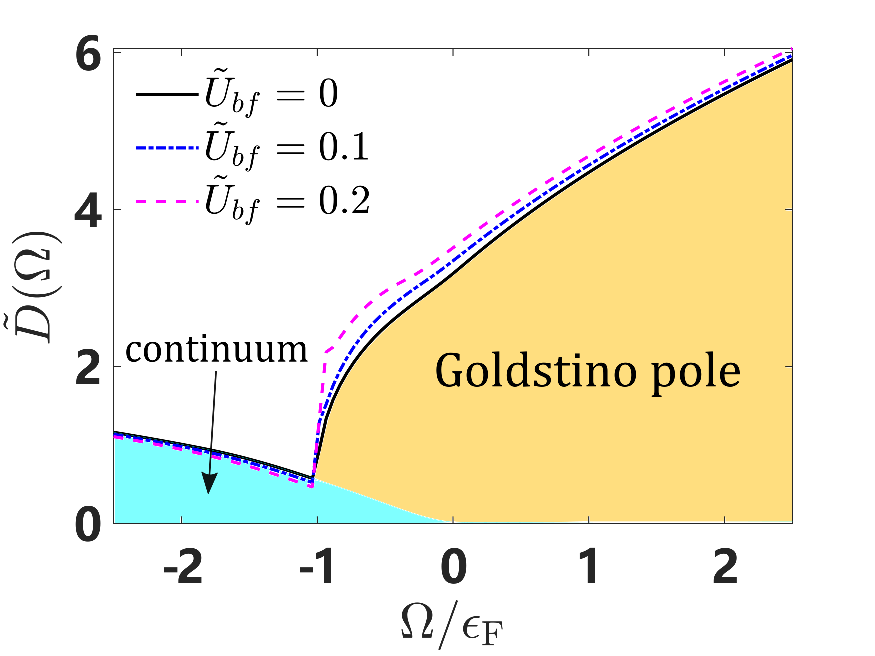}
    \caption{Goldstino density of states with different particle number densities in a $^{173}$Yb-$^{174}$Yb mixture, where $\tilde{D}(\Omega)=\epsilon_{\rm F}D(\Omega)/(6N^2_f)$. The boson-to-fermion particle number ratio is set to be $N_b/N_f=2$.
    }\label{DOS}
\end{figure}
\begin{figure}[t]
    \centering
    \includegraphics[width=8cm]{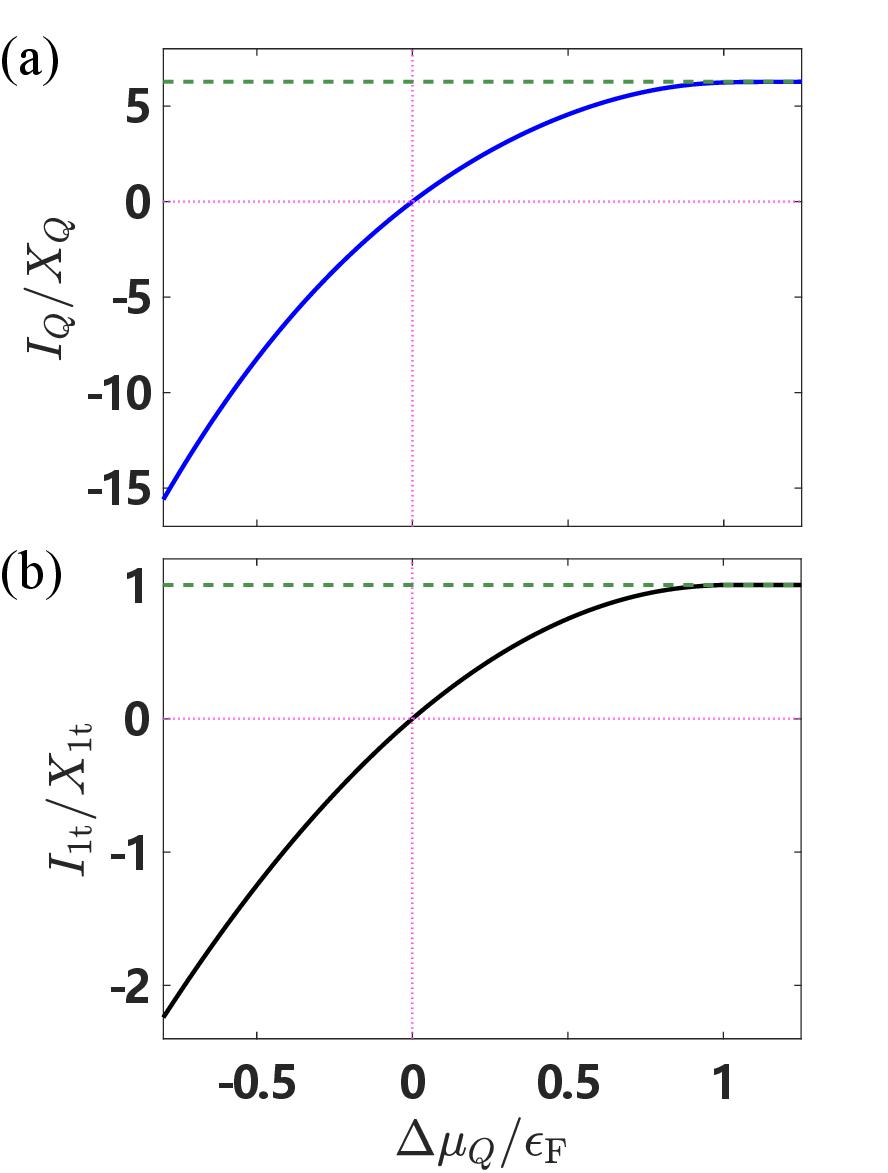}
    \caption{(a) Bias dependence of the goldstino tunneling current in the $^{173}$Yb-$^{174}$Yb mixture. (b) Bias dependence of quasiparticle current $I_{\rm 1t}$.
    Here, $X_{Q}=9\mathcal{G}_{\bm{Q}}^2 (2N_{f,{\rm L}})^4/(\epsilon_{\rm F}\pi)$ and $X_{\rm 1t}=9\pi \mathcal{T}_f^2N_{f,{\rm L}}^2/(4\epsilon_{\rm F})$ are normalizing factors of $I_{Q}$ and $I_{\rm 1t}$, respectively. We fix the Fermi energy for the left reservoir to be $\epsilon_{\rm F}$. The ratio of particle numbers is fixed to be $N_{b,i}/N_{f,i}=2$ and in both reservoirs the dimensionless interaction strength is taken to be $\tilde{U}_{bf}=0.2$.
    The horizontal dotted lines in each panel represent the correspoding values at large $\Delta\mu_Q\rightarrow\infty$.
    }\label{I-V}
\end{figure}

Defining the dimensionless interaction strength $\tilde{U}_{bf}=2U_{bf}N_f/\epsilon_{\rm F}=8a_{bf}k_{\rm F}/(3\pi)$ and $\tilde{U}_{bb}=2U_{bb}N_f/\epsilon_{\rm F}=8a_{bb}k_{\rm F}/(3\pi)$ with $\epsilon_{\rm F}$ and $k_{\rm F}$ denoting the Fermi energy and Fermi momentum of the fermions, we calculated the goldstino spectra in a $^{173}$Yb-$^{174}$Yb mixture, where the boson-fermion and boson-boson scattering lengths are precisely determined by experiments as $a_{bf}=138.49 a_0$ and $a_{bb}=104.72 a_0$~\cite{PhysRevA.82.011608} ($a_0$ is the Bohr radius). Since the scattering lengths can not be tuned via magnetic Feshbach resonances in a $^{173}$Yb-$^{174}$Yb mixture, $\tilde{U}_{bf}$ can only be changed by $k_{\rm F}$, namely, the density $N_f$. For a certain value of $\tilde{U}_{bf}$, the corresponding density is $N_f\simeq 1.04\times10^{-8}\tilde{U}^3_{bf}/a_0^3$.
Figure~\ref{Imchi} shows the spectra with different particle densities, where particle numbers of two components are taken to be $N_b=2N_f$. The goldstino poles denoted by a sharp peak arises from the condensate of bosons, and yields a nonzero energy gap, which is caused by the explicit supersymmetry breaking associated with the chemical potential bias between fermions and bosons.
The sharp-peaked structure indicates its long lifetime collective excitation, which is reminiscent of the bulk dissipationless flow observed in a strongly-interacting $^{23}$Na - $^{40}$K mixture~\cite{yan2023dissipationless}.
As the interaction strength increases, the pole gradually separates from the continuum, and completely leaves away from the continuum when $\tilde{U}_{bf}\simeq 0.3$.
Note that in the repulsively interacting regime, bosons and fermions tend to be spatially separated beyond a critical interaction strength~\cite{PhysRevA.61.053605,PhysRevA.66.063604}. 
Such phase separation has recently been studied in mass-imbalanced mixtures, where the phase separation occurs at the repulsive branch when tuning the $a_{bf}$ from a small positive value to resonance~\cite{PhysRevLett.120.243403,PhysRevLett.131.083003,PhysRevLett.132.033401}. 
This process is similar to the Stoner ferromagnetic phase transition in a two-spin-component Fermi gases~\cite{Stoner,science.1177112}. To keep the stability of mixture against the phase separation, we focus on the weak coupling regime ($\tilde{U}_{bf}<0.4$).

It is useful to define the goldstino density of state (DOS) as
\begin{align}\label{Dos}
    D(\Omega) =\sum_{\bm{p}}{\rm Im}\chi_{\bm{p}}(\Omega).
\end{align}
According to Eqs.~(\ref{Lindhard1}) and (\ref{RPA}), it consists of contributions from the poles and continuum part, while that of pole has a nonzero value only when $\Omega+\epsilon_{\rm F}>0$. We focus on the density below the critical point indicated by the goldstino spectrum and at extremely low temperature such that $N^0_b\rightarrow N_b$ and Eq.~(\ref{Pic0}) holds.
Figure~\ref{DOS} shows the normalized goldstino DOS $\tilde{D}(\Omega)$ in a $^{173}$Yb-$^{174}$Yb mixture with different particle number densities and the fixed particle number ratio $N_{b,i}/N_{f,i}=2$. We can see that when $\Omega>-\epsilon_{\rm F}$, the contribution of poles arises and dominants the DOS.  
\begin{figure}[t]
    \centering
    \includegraphics[width=8.6cm]{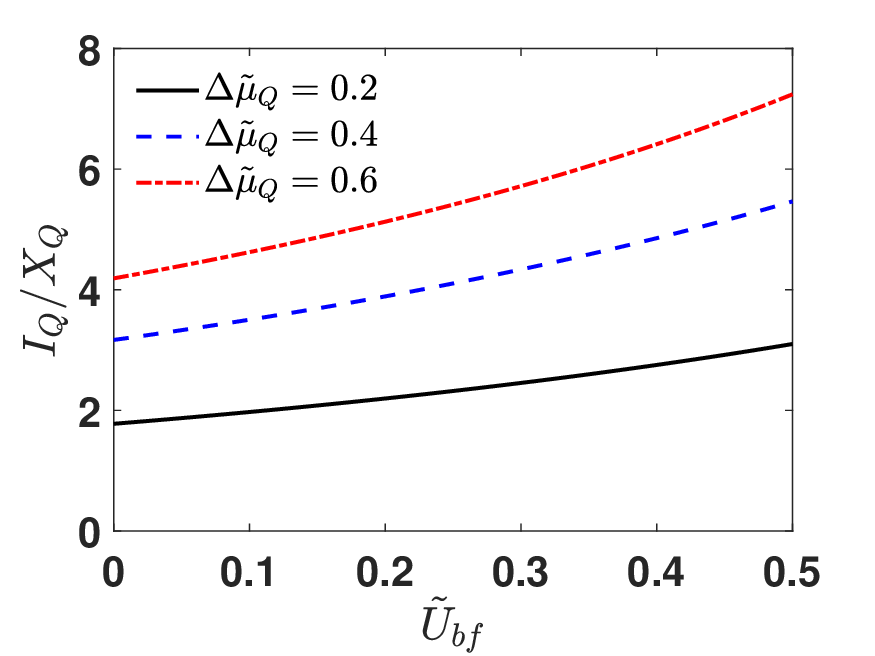}
    \caption{Current-interaction features for the goldstino tunneling with different chemical potential bias. The boson-boson interaction strength is fixed to be $\tilde{U}_{bb}=0.15$, while $\tilde{U}_{bf}$ is tuned from $0$ to $0.5$.
    }\label{I-U}
\end{figure}

\subsection{Supercharge tunneling current}

Using the DOS given by Eq.~\eqref{Dos}, one can rewrite the goldstino tunneling current Eq.~(\ref{Isc}) as 
\begin{align}\label{Isc2}
    I_{Q}=8\mathcal{G}_{Q}^{2} &\int \frac{d \Omega}{2 \pi} D_{\rm L}(\Omega-\Delta\mu_{Q}) D_{\rm R}(\Omega)\cr
    &\times\left[{f}_f(\Omega-\Delta\mu_{Q})-{f}_f(\Omega)\right].
\end{align}
The bias dependence of $I_Q$ is shown in Fig.~\ref{I-V}(a), where we consider the interaction parameters in the $^{173}$Yb-$^{174}$Yb mixture with the dimensionless interaction strength $\tilde{U}_{bf}=0.2$.
We notice that $I_{Q}$ increases monotonically with the chemical potential bias within the interval $0<\Delta\mu_{Q}/\epsilon_{\rm F}<1$ but approaches a constant $I_{Q}/X_{Q}\simeq 6.3$ at $\Delta\mu_Q/\epsilon_{\rm F}\geq 1$.  This behavior indicates a vanishing differential conductance
which is caused by the energy shift acting on the goldstino spectra by the large bias.
Moreover, one can find an asymmetric current-bias characteristics $I_{Q}(\Delta\mu_Q)\neq-I_{Q}(-\Delta\mu_Q)$.
These features are distinct from the magnon-tunneling current~\cite{PhysRevB.108.155303,PhysRevApplied.21.L031001} due to the statistical difference of the fermionic and bosonic Nambu-Goldstone modes.
Indeed, the qualitative behavior of $I_{Q}$ is similar to that of the fermionic quasiparticle current $I_{\rm 1t}$ in Fig.~\ref{I-V}(b), where the analytical expression of $I_{\rm 1t}$ is shown in Appendix~\ref{appendixB}.

From the expression of the chemical potentials at zero temperature, we find that $\Delta\mu_{Q}=\Delta\mu_{f}$ in the case of balanced bosonic chemical potential.
Accordingly, the goldstino pole energy is shifted by the change of the Fermi energy $\epsilon_{{\rm F},i}=\mu_{f,i}-U_{bb}N_{b,i}$ in each bulk reservoir.
 In this study, since we fix the Fermi energy in the left reservoir (i.e., $\epsilon_{\rm F,L}\equiv\epsilon_{\rm F}$), a nonzero $\Delta\mu_{f}$ leads to shift of $\mu_{f,{\rm R}}=\mu_{f,{\rm L}}-\Delta\mu_{f}$ and thus the goldstino DOS in the right reservoir.
Moreover, the energy shift $-\Delta\mu_{Q}$ is included in $D_{\rm L}$ in Eq.~(\ref{Isc2}), which leads to the same energy shift of the DOS in the left reservoir.
Therefore, the energy shift of goldstino poles in both reservoirs is equal to each other even as the bias increases. On the other hand, at extremely low temperature, the difference between two distribution functions in Eq.~(\ref{Isc2}) yields a nonzero interval $0<\Omega<\Delta\mu_{Q}$. 
As a result, when the value of $\Delta\mu_{Q}/\epsilon_{\rm F}$ exceeds $1$,
$I_{Q}$ saturates at a finite value due to the Fermi statistics of goldstino.
Such feature is similar to that of the quasiparticle tunneling current shown in Fig.~\ref{I-V}(b), while the zero conductance of the later one is due to the zero chemical potential bias for bosons that we choose. In spite of the similar behavior of these two tunneling signals, they may be distinguished through the current shot noise, which is proportional to the charge of the carriers and the average current~\cite{jehl2000detection,PhysRevLett.84.3398}. The noise-to-current ratio is found to be a straightforward probe of tunneling channel~\cite{10.1093/pnasnexus/pgad045,PhysRevApplied.21.L031001}, while it increases from $1$ to $2$ as the interaction strength is enhanced, indicating a crossover from one-body to multi-particle tunneling in an itinerant Fermi gas. Therefore it is possible to find the dominant goldstino tunneling by measuring the noise and adjusting the interaction.

While the results presented above are specific to the $^{173}$Yb-$^{174}$Yb mixture, our methodology is applicable to other Bose-Fermi mixtures with small mass imbalances, such as $^6$Li-$^7$Li~\cite{doi:10.1126/science.1255380,PhysRevLett.118.103403,Ikemachi_2017}, $^{39}$K-$^{40}$K~\cite{PhysRevA.78.012503}, $^{40}$K-$^{41}$K~\cite{PhysRevA.84.011601}, $^{84}$Sr-$^{87}$Sr~\cite{PhysRevA.82.011608}, $^{87}$Rb-$^{87}$Sr~\cite{barbe2018observation}, and $^{161}$Dy-$^{162}$Dy~\cite{PhysRevLett.103.085301}, provided they are below their respective Bose-Einstein condensation (BEC) temperatures. Additionally, in some of these mixtures, interactions between bosons or fermions can be adjusted by manipulating scattering lengths through magnetic Feshbach resonances~\cite{PhysRevA.84.011601,barbe2018observation,D'Errico_2007,PhysRevLett.102.090402,PhysRevA.79.031602}. This flexibility allows for exploring goldstino spectral features and its transport properties across various interaction strength regimes.

Motivated by this, we investigate the interaction dependence of the goldstino tunneling current, as shown in Fig.~{\ref{I-U}}, where we fix the boson-boson interaction strength at $\tilde{U}_{bb}=0.15$ and tune the boson-fermion interaction through a Feshbach resonance between species. We should note that although the beyond-mean-field theories demonstrate a $1/k^4$ tail for single-particle momentum distribution in short-range interacting regime~\cite{TAN20082952,TAN20082971,TAN20082987}, indicating that the quasiparticle tunneling current should be modified, we only consider the weak-coupling regime where the mean-field approximation can be employed and the $1/k^4$ momentum tail can be neglected. On the other hand, we can see from its expression (Eq.~(\ref{I1t1})) that while at finite temperature the quasiparticle tunneling would show dependence on the interaction due to the mean-field energy corrections, at extremely low temperatures the corrections are offset by the zero-temperature chemical potential, making $I_{\rm 1t}$ interaction-independent. In contrast, the goldstino tunneling current is enhanced as the interaction strength increases even in zero temperature, providing a potential way to distinguish between these two signals. We should also note that the ratio between two normalizing factors is given by ${X_Q}/{X_{\rm 1t}}={16\tilde{U}_{bf}^2\gamma^2}/{\pi^2}$, where $\gamma=\frac{\epsilon_{\rm F}}{U_{bf}}\frac{\mathcal{G}_{\bm{Q}}}{\mathcal{T}_f}$ is a dimensionless ratio with the order of magnitude $\gamma \sim 1$~\cite{PhysRevApplied.21.L031001}. Then combining with Fig.~\ref{I-V}, we find that when $\tilde{U}_{bf}$ reaches around $0.3$, $I_Q$ would have the same order of magnitude as $I_{\rm 1t}$, which means one can easily distinguish them with a relatively low accuracy of measurement.

Moreover, the shape of the potential barrier (e.g., height and width) can also influence $\mathcal{T}_{f}$ and $\mathcal{G}_{Q}$ through the modification of the transmission coefficients. 
While both one-body and two-body tunneling strengths decrease as the height and width increase, the latter is shown to be more sensitive to the barrier~\cite{PhysRevApplied.21.L031001}.
Such different dependencies on the shape of the potential barrier may also aid in distinguishing the goldstino tunneling signal from the quasiparticle one. 

\section{Summary and perspectives}\label{Summary}

In this study, we have theoretically examined the tunneling transport induced by a chemical potential bias in a Bose-Fermi mixture, which can be a promising route
to detect the goldstino excitation associated with the broken supersymmetry in ultracold atomic experiments. 

By calculating the goldstino spectrum and employing the Schwinger-Keldysh approach up to the leading order, we explore the supersymmetry-like tunneling current through a two-terminal junction induced by a chemical potential bias between two reservoirs. 
The goldstino mode exhibits a chemical potential of $\mu_f-\mu_b$ and is found to contribute to the tunneling process. 
In spite of the similar bias dependence of the quasiparticle and goldstino tunneling currents, in Bose-Fermi mixtures allowing interaction tuning through Feshbach resonances, they may be distinguished through the interaction dependence. 

Although the gapped goldstino excitation and its spectral properties have not been directly observed, our work suggests a potential avenue for probing them through tunneling transport, which can be conducted and detected in laboratories. Once the supercharge current is determined and agree with the results of theoretical analysis, it would serve as a strong evidence for the existence of the goldstino and supersymmetry in such Bose-Fermi mixture systems.  
Our approach can also be applied to other condensed-matter systems exhibiting the supersymmetric properties~\cite{PhysRevD.99.045002,marra20221d,miura2023supersymmetry}. 

\begin{acknowledgements}
H. T. thanks Yoshimasa Hidaka and Daisuke Satow for useful discussions.
T. Z. and Y. G. were supported by the RIKEN Junior Research Associate Program.
H. T. acknowledges the JSPS Grants-in-Aid for Scientific Research under Grants No.~22H01158, and No.~22K13981.
H. L. acknowledges the JSPS Grant-in-Aid for Scientific Research (S) under Grant No.~20H05648 and the RIKEN Pioneering Project: Evolution of Matter in the Universe.

T. Z. and Y. G. contributed equally to this paper and should be considered as co-first authors.

\end{acknowledgements}

\appendix

\section{Calculation of goldstino propagator}\label{appendixA}

In this appendix, we give the details in calculating the Goldstino spectra.
Notice that the contribution of continuum in Eq.~(\ref{Lindhard1}) can be rewritten as 
\begin{align}
     \Pi^{\rm c}_{\bm{p}}(\Omega)=-\int\frac{d^3\bm{k}}{(2\pi)^3}\bigg[&\frac{{f}_f(\xi_{\bm{k},f})}{\Omega+i\delta-\xi_{\bm{k},f}+\xi_{\bm{k}-\bm{p},b}}\cr
     +&\frac{{f}_b(\xi_{\bm{k},b})}{\Omega+i\delta-\xi_{\bm{k}+\bm{p},f}+\xi_{\bm{k},b}}\bigg].
\end{align}
We calculate the integral in a spherical coordinate system, where, according to the symmetry, we can take the vector $\bm{p}$ coincident with $z$ axis. Taking $k_{\rm F}$ and $\epsilon_{\rm F}$ as the Fermi momentum and Fermi energy of the fermions, we define $\tilde{k}=k/k_{\rm F}$, $\tilde{\Omega}=\Omega/\epsilon_{\rm F}$, and $\tilde{\mu}_{Q}=\mu_{Q}/\epsilon_{\rm F}$. Then after conducting the angular integral, we have 
\begin{align}
    \Pi^{\rm c}_{\bm{p}}(\Omega)&=\frac{k_{\rm F}^3}{8\pi^2\epsilon_{\rm F}\tilde{p}}\int d\tilde{k}\tilde{k}\cr
    \bigg[&{f}_f(\xi_{\bm{k},f})\ln\bigg(\frac{\tilde{\Omega}+i\delta-2\tilde{k}\tilde{p}+\tilde{p}^2+\tilde{\mu}'_{Q}}{\tilde{\Omega}+i\delta+2\tilde{k}\tilde{p}+\tilde{p}^2+\tilde{\mu}'_{Q}}\bigg)\nonumber\\
    +&{f}_b(\xi_{\bm{k},b})\ln\bigg(\frac{\tilde{\Omega}+i\delta-2\tilde{k}\tilde{p}-\tilde{p}^2+\tilde{\mu}'_{Q}}{\tilde{\Omega}+i\delta+2\tilde{k}\tilde{p}-\tilde{p}^2+\tilde{\mu}'_{Q}}\bigg)\bigg],
\end{align}
where $\mu'_{Q}=\mu_{Q}-(\Sigma_f-\Sigma_b)=\epsilon_{\rm F}+U_{bb}N_b$.
Due to the infinitesimally small number $\delta$, the formulas inside the parentheses can be expressed as 
\begin{align}
    &\frac{\tilde{\Omega}+i\delta-2\tilde{k}\tilde{p}\pm\tilde{p}^2+\tilde{\mu}'_{Q}}{\tilde{\Omega}+i\delta+2\tilde{k}\tilde{p}\pm\tilde{p}^2+\tilde{\mu}'_{Q}}\cr
    =&~\frac{\big|(\tilde{\Omega}\pm\tilde{p}^2+\tilde{\mu}'_{Q})^2-4\tilde{k}^2\tilde{p}^2\big|}{(\tilde{\Omega}+2\tilde{k}\tilde{p}\pm\tilde{p}^2+\tilde{\mu}'_{Q})^2}\cr
    & \times \exp\bigg\{i\arctan\bigg[\frac{4\tilde{k}\tilde{p}\delta}{(\tilde{\Omega}\pm\tilde{p}^2+\tilde{\mu}'_{Q})^2-4\tilde{k}^2\tilde{p}^2}\bigg]\bigg\}\cr
    =&~A^\pm_{\tilde{\bm{p}}}(\tilde{\Omega},\tilde{k})e^{i\tilde{\theta}^\pm_{\bm{p}}(\tilde{\Omega},\tilde{k})}.
\end{align}
Then we obtain the real part of the one-loop Goldstino propagator:
\begin{align}
    \operatorname{Re}\Pi^{\rm c}_{\bm{p}}(\Omega)=\frac{3N_f}{4\epsilon_{\rm F}\tilde{p}}\int &d\tilde{k}\,\tilde{k}\big\{{f}_f(\xi_{\bm{k},f})\ln\big[A^+_{\tilde{\bm{p}}}(\tilde{\Omega},\tilde{k})\big]\cr
    &+{f}_b(\xi_{\bm{k},b})\ln\big[A^-_{\tilde{\bm{p}}}(\tilde{\Omega},\tilde{k})\big]\big\},
\end{align}
where the fermionic density reads $N_f=k_{\rm F}^3/6\pi^2$.
For the imaginary part of $\Pi$, according to the identity $\frac{1}{\Gamma+i\delta}=\mathcal{P}\frac{1}{\Gamma}-i\pi\delta(\Gamma)$, we have $\operatorname{Im}\Pi_{\bm{p}}=\operatorname{Im}\Pi^{\rm p}_{\bm{p}}+\operatorname{Im}\Pi^{\rm c}_{\bm{p}}$, where 
\begin{align}\label{ImPip}
    \operatorname{Im}\Pi^{\rm p}_{\bm{p}}(\Omega)=\pi N^0_b\delta(\Omega+\epsilon_{\rm F}-\bm{p}^2/2m)
\end{align}
and
\begin{align}\label{ImPic}
    \operatorname{Im}\Pi^{\rm c}_{\bm{p}}(\Omega)&=\pi\int\frac{d^3\bm{k}}{(2\pi)^3}\big[{f}_f(\xi_{\bm{k}+\bm{p},f})+{f}_b(\xi_{\bm{k},b})\big]\cr
    &\delta\big(\Omega-(\bm{k}+\bm{p})^2/2m+\bm{k}^2/2m+\mu'_{Q}\big).
\end{align}
By expressing the integral over parameters in the spherical coordinates and performing a variable conversion, where $\cos{\theta}$ is replaced by $q=|\bm{k}+\bm{p}|$, and introducing the parameters such as $\tilde{k}$ and $\tilde{\Omega}$ that are normalized by $k_{\rm F}$ and $\epsilon_{\rm F}$, the equation above can be reformulated as
\begin{align}
    \operatorname{Im}\Pi^{\rm c}_{\bm{p}}(\Omega)&=\frac{k_{\rm F}^3}{4\pi\epsilon_{\rm F}\tilde{p}}\int\tilde{k}d\tilde{k}\int_{|\tilde{k}-\tilde{p}|}^{\tilde{k}+\tilde{p}}\tilde{q}d\tilde{q}\nonumber\\
    &\big[{f}_f(\xi_{\bm{q},f})+{f}_b(\xi_{\bm{k},b})\big]\delta\big(\tilde{\Omega}-\tilde{q}^2+\tilde{k}^2+\tilde{\mu}'_{Q}\big).
\end{align}
By using the property of delta function: $\delta[f(x)]=\delta(x-x_0)/|f'(x_0)|$ with $f(x_0)=0$, we have
\begin{align}
    \operatorname{Im}\Pi^{\rm c}_{\bm{p}}(\Omega)=&~\frac{k_{\rm F}^3}{8\pi\epsilon_{\rm F}\tilde{p}}\int\tilde{k}d\tilde{k}\int_{|\tilde{k}-\tilde{p}|}^{\tilde{k}+\tilde{p}}d\tilde{q}\nonumber\\
    &\big[{f}_f(\xi_{\bm{q},f})+{f}_b(\xi_{\bm{k},b})\big]\delta(\tilde{q}-\tilde{q}_0),
\end{align}
where $\tilde{q}_0=\sqrt{\tilde{\Omega}+\tilde{k}^2+\tilde{\mu}'_{Q}}$. Meanwhile, for the integral to be nonzero, $\tilde{q}_0$ must satisfy the inequality $|\tilde{k}-\tilde{p}|\leq \tilde{q}_0\leq \tilde{k}+\tilde{p}$, leading to a lower limit for the integral over $k$ as
\begin{align}
    \tilde{k}\geq\frac{1}{2}\bigg|\frac{\tilde{\Omega}+\tilde{\mu}'_{Q}}{\tilde{p}}-\tilde{p}\bigg|.
\end{align}
Therefore, we obtain the expression for the imaginary part of the Lindhard function as given by
\begin{subequations}
    \begin{align}
    \operatorname{Im}\Pi^{\rm p}_{\bm{p}}(\Omega)=\pi N^0_b\delta(\Omega+\epsilon_{\rm F}-\bm{p}^2/2m),
\end{align}
\begin{align}
    \operatorname{Im}\Pi^{\rm c}_{\bm{p}}(\Omega)=\frac{3\pi N_f}{4\epsilon_{\rm F}\tilde{p}}\int_{\alpha}^\infty\tilde{k}d\tilde{k}\big[{f}_f(\xi_{q_0,f})+{f}_b(\xi_{k,b})\big],
\end{align}
\end{subequations}
where $\alpha=\frac{1}{2}\big|\frac{\tilde{\Omega}+\tilde{\mu}'_{Q}}{\tilde{p}}-\tilde{p}\big|$.

Introducing the dimensionless interaction strength $\tilde{U}_{ij}=2U_{ij}N_f/\epsilon_{\rm F}=8a_{ij}k_{\rm F}/(3\pi)$, the Goldstino spectral function within the RPA can be computed as
\begin{align}\label{ImchiRPA}
    \operatorname{Im}\tilde{\chi}_{\bm{p}}(\Omega)=\frac{\operatorname{Im}
    \tilde{\Pi}_{\bm{p}}(\Omega)}{(1+\tilde{U}_{bf}\operatorname{Re}\tilde{\Pi}^{\rm c}_{\bm{p}}(\Omega))^2+(\tilde{U}_{bf}\operatorname{Im}\tilde{\Pi}^{\rm c}_{\bm{p}}(\Omega))^2},
\end{align}
where $\tilde{\chi}_{\bm{p}}=\chi_{\bm{p}}\epsilon_{\rm F}/(2N_f)$ and $\tilde{\Pi}_{\bm{p},i}=\Pi_{\bm{p}}\epsilon_{\rm F}/(2N_f)$. 
The Goldstino DOS can be calculated by changing the discrete summation in Eq.~(\ref{Dos}) into the integral over parameters in a spherical coordinate as
\begin{widetext}
    \begin{align}
    \tilde{D}(\Omega)=\int d\tilde{q}\tilde{q}^2\operatorname{Im}\tilde{\chi}^{\rm ret.}_{\bm{q}}(\Omega)
    =&~\Theta(\tilde{\Omega}+1)\frac{\pi N_b}{2N_f}\frac{\tilde{q}_0}{(1+\tilde{U}_{bf}\operatorname{Re}\tilde{\Pi}^{\rm c}_{q_0}(\Omega))^2+(\tilde{U}_{bf}\operatorname{Im}\tilde{\Pi}^{\rm c}_{q_0}(\Omega))^2}\nonumber\\
    &+\int d\tilde{q}\tilde{q}^2 \frac{\operatorname{Im}
    \tilde{\Pi}^{\rm c}_{\bm{q}}(\Omega)}{(1+\tilde{U}_{bf}\operatorname{Re}\tilde{\Pi}^{\rm c}_{\bm{q}}(\Omega))^2+(\tilde{U}_{bf}\operatorname{Im}\tilde{\Pi}^{\rm c}_{\bm{q}}(\Omega))^2},
\end{align}
\end{widetext}
where $q_0=\sqrt{2m(\Omega+\epsilon_{\rm F})}$. The first term on the right hand side describes the contribution of the pole with a gap $\Delta_{\rm p}=-\epsilon_{\rm F}$, while the second one corresponds to that of the continuum.
\begin{figure}[t]
    \centering
    \includegraphics[width=8.6cm]{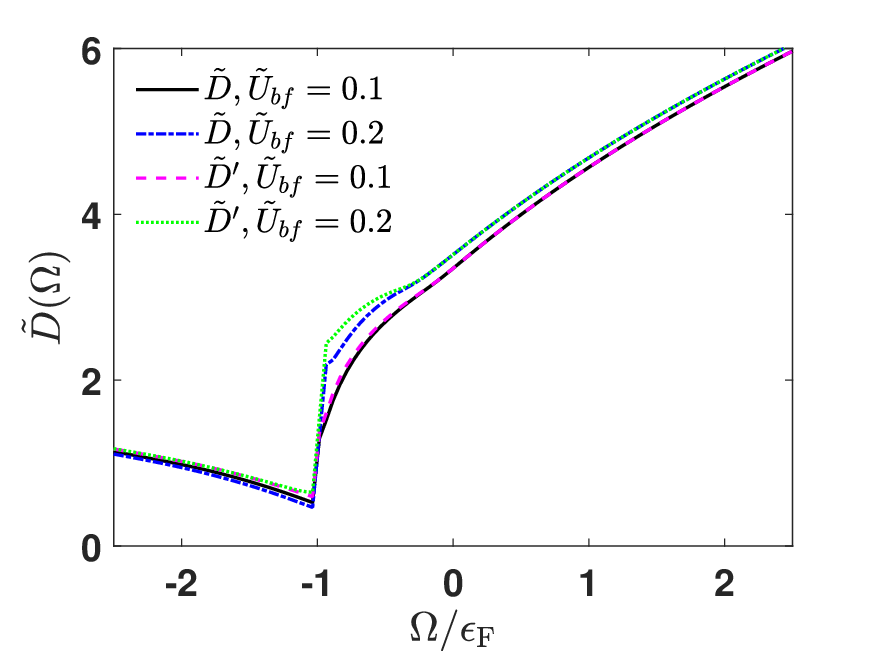}
    \caption{Goldstino density of states with different particle number densities in a $^{173}$Yb-$^{174}$Yb mixture, where we compared the results taking $\operatorname{Re}\Pi^{\rm p}$ into account ($\tilde{D}^\prime$) with the results not taking $\operatorname{Re}\Pi^{\rm p}$ into account ($\tilde{D}$). The boson-to-fermion particle number ratio is set to be $N_b/N_f=2$.
    }\label{DOS1}
\end{figure}

Notice that in the denominator of Eq.~\eqref{ImchiRPA}, we omit the contribution of $\Pi^{\rm p}$ whose spectrum is a delta function which does not affect the result of later integral calculations. 
We also omit the contribution of  $\operatorname{Re}\Pi^{\rm p}$ which has a negligible effect to the Goldstino density states in our weakly interacting regime, as shown in Fig.~\ref{DOS1}.

\section{Calculation of quasiparticle tunneling current}\label{appendixB}

In this appendix, we provide the analytical derivations of quasiparticle tunneling current $I_{\rm 1t}$ for understanding the asymmetric characteristic clearly.

The Green's function of single fermion at zero temperature is given by 
\begin{align}
    G_{f,\bm{k}}(\omega)=\frac{1}{\omega-\xi_{\bm{k},f}
    +i\delta},
\end{align}
where $\xi_{\bm{k},f}=\bm{k}^2/2m-\epsilon_{\rm F}$ is the single-particle energy for fermions in the grand-canonical ensemble after taking into account the self-energy correction. Its imaginary part reads 
\begin{align}
    \operatorname{Im}G_{f,\bm{k}}(\omega)=-\pi\delta(\omega-\xi_{\bm{k},f}).
\end{align}
According to Eq.~(\ref{I1t1}), the quasiparticle current can then be rewritten as 
\begin{align}
    I_{\rm 1t}=&~4\mathcal{T}_{f}^2\int\frac{d\omega}{2\pi}\int\frac{d^3\bm{p}}{(2\pi)^3}\int\frac{d^3\bm{q}}{(2\pi)^3}\pi^2\delta(\omega-\xi_{\bm{p},f,{\rm L}}-\Delta\mu_f)\cr
    &\times\delta(\omega-\xi_{\bm{q},f,{\rm R}})[{f}_f(\omega-\Delta\mu_f)-{f}_f(\Omega)].
\end{align}
In ultracold regimes, the Fermi distribution function can be approximately regarded as ${f}_f(\omega)\simeq\theta(-\omega)$. Fixing the Fermi energy of the left reservoir to be $\epsilon_{\rm F}$, we have 
\begin{align}
    I_{\rm 1t}\simeq&~\frac{\mathcal{T}_f^2k_{\rm F}^6}{8\pi^3\epsilon_{\rm F}}\int d\tilde{\omega}\sqrt{\tilde{\omega}+1-\Delta\tilde{\mu}_f}\sqrt{\tilde{\omega}+1-\Delta\tilde{\mu}_f}\nonumber\\
    &\times\theta(\tilde{\omega}+1-\Delta\tilde{\mu}_f)[\theta(\Delta\tilde{\mu_f}-\tilde{\omega})-\theta(-\tilde{\omega})]\nonumber\\
    =&~\frac{\mathcal{T}_f^2k_{\rm F}^6}{8\pi^3\epsilon_{\rm F}}\int_0^{\Delta\tilde{\mu}_f}d\tilde{\omega}(\tilde{\omega}+1-\Delta\tilde{\mu}_f)\theta(\tilde{\omega}+1-\Delta\tilde{\mu}_f)\nonumber\\
    =&~X_{\rm 1t}\bigg\{\Big[(\tilde{\omega}+1-\Delta\tilde{\mu}_f)^2\Big]_0^{\Delta\tilde{\mu}_f}\theta(1-\Delta\tilde{\mu}_f)\nonumber\\
    &+\Big[(\tilde{\omega}+1-\Delta\tilde{\mu}_f)^2\Big]_{\Delta\tilde{\mu}_f-1}^{\Delta\tilde{\mu}_f}\theta(\Delta\tilde{\mu}_f-1)\bigg\}\nonumber\\
    =&~X_{\rm 1t}\big[1-(1-\Delta\tilde{\mu}_f)^2\theta(1-\Delta\tilde{\mu}_f)\big],
\end{align}
where $\Delta\tilde{\mu}_f=\Delta\mu_f/\epsilon_{\rm F}$. In the present case with the balanced bosonic chemical potential in the two reservoirs, one can find $\Delta\mu_f=\Delta\mu_Q$.

From the derivation above, we can see that the asymmetric characteristic arises from the fixed Fermi energy in the left reservoir, while the vanishing differential conductance is due to the disappearance of the Fermi energy in the right reservoir.

\bibliography{ref.bib}

\end{document}